\documentclass[prl,superscriptaddress,twocolumn,showpacs]{revtex4}     
\usepackage{times}

\usepackage{./aas_macros}
\usepackage{graphicx}
\graphicspath{{images/}}
\usepackage{amssymb}
\usepackage{textcomp}
\usepackage{color}

\begin{document}

\title{Imprints of superfluidity on magneto-elastic QPOs of SGRs}

\author{Michael Gabler} 
\email{michael.gabler@uv.es}
\affiliation{Departamento de Astronom\'{\i}a y Astrof\'{\i}sica,
  Universitat de Val\`encia, 46100 Burjassot (Valencia), Spain}
\affiliation{Max-Planck-Institut f\"ur Astrophysik,
  Karl-Schwarzschild-Str.~1, 85741 Garching, Germany}

\author{Pablo Cerd\'a-Dur\'an}
\affiliation{Departamento de Astronom\'{\i}a y Astrof\'{\i}sica,
  Universitat de Val\`encia, 46100 Burjassot (Valencia), Spain}

\author{Nikolaos Stergioulas} 
\affiliation{Department of Physics, Aristotle University of Thessaloniki,
  Thessaloniki 54124, Greece}

\author{Jos\'e A.~Font}
\affiliation{Departamento de Astronom\'{\i}a y Astrof\'{\i}sica,
  Universitat de Val\`encia, 46100 Burjassot (Valencia), Spain}

\author{Ewald M\"uller}
\affiliation{Max-Planck-Institut f\"ur Astrophysik,
  Karl-Schwarzschild-Str.~1, 85741 Garching, Germany}
\date{\today}
\begin{abstract}
Our numerical simulations show that axisymmetric, torsional,
magneto-elastic oscillations of magnetars with a superfluid core
can explain the whole range of observed quasi-periodic oscillations
(QPOs) in the giant flares of soft gamma-ray repeaters. There exist
constant phase, magneto-elastic QPOs at both low ($f<150$Hz) and high
frequencies ($f>500$Hz), in full agreement with observations. The
range of magnetic field strengths required to match the observed QPO
frequencies agrees with that from spin-down estimates. These results
strongly suggest that neutrons in magnetar cores are superfluid.
\end{abstract}
\keywords{Neutron stars; Oscillations stellar;
Stellar seismology; Magnetohydrodynamics in astrophysics;
Superfluidity hydrodynamic aspects of}
\pacs{97.60.Jd, 97.10.Sj, 97.10.Sj, 95.30.Qd, 47.37.+q}
\maketitle
%
%

Neutron stars are perfect astrophysical laboratories to study the
equation of state (EoS) of matter at supra-nuclear densities, i.e., at
conditions impossible to replicate on Earth.  Giant flares of Soft
Gamma-ray Repeaters (SGRs) are very promising events that can be used
to obtain information about the structure of neutron stars, since it
is believed that their source are highly magnetized neutron stars
(magnetars) \cite{Duncan1992} suffering a global rearrangement of the
magnetic field, and possibly involving a fracture of the solid
crust. In the X-ray light curves of two of the three giant flares
detected so far, SGR 1806-20 and SGR 1900+14, a number of
Quasi-Periodic Oscillations (QPOs) have been observed
\cite{Israel2005, Watts2007}. This may have been the first detection
of neutron star oscillations, which provide a possibility for studying
such compact stars through asteroseismology. The observed frequencies
consist of two categories, {\it low frequency QPOs} between a few tens
of Hz and up to $150\,$Hz observed in both events, and {\it high
  frequency QPOs} above $500\,$Hz, which are only observed in the 2004
giant flare. Some QPO frequencies roughly match those of discrete
crustal shear modes in non-magnetized stars, namely $n=0$ torsional
modes (nodeless in the radial direction) for the low frequency QPOs
and $n\geq1$ modes for the high frequency QPOs (see \cite{Duncan1998,
  Strohmayer2005, Piro2005, Sotani2007, Samuelsson2007} and references
therein).  However, these crustal modes are quickly damped by the
magnetic field in the core \cite{Levin2007, Gabler2011letter,
  Gabler2012, Colaiuda2011, vanHoven2011, vanHoven2012, Gabler2013}.

On the other hand, torsional Alfv\'en oscillations (fundamental mode
$\sim30\,$Hz), i.e. QPOs trapped at turning-points or edges of the
Alfv\'en continuum of the highly magnetized core, can also have
frequencies similar to those of the observed QPOs for magnetar field
strengths of order $B\sim10^{15}\,$G, with the additional attractive
feature of overtones appearing at near-integer ratios
\cite{Sotani2008, Cerda2009, Colaiuda2009}.

The Alfv\'en QPO model extended to magneto-elastic QPOs and different
types of magnetic fields \cite{Gabler2012, Gabler2013} explains the
observed low frequency QPOs as excitations of a fundamental
turning-point QPO and of several overtones.  However, the observation
of high frequency QPOs poses a problem for this model, because the
first overtone ($n=1$) crustal shear mode is quickly absorbed into the
Alfv\'en continuum \cite{Gabler2012,vanHoven2012} and because there is
no known mechanism to excite a specific high-order overtone of the
turning-point magneto-elastic QPOs with the appropriate frequencies.
A model explaining both low- and high-frequency QPOs would thus be a
significant step towards a better understanding of neutron star
interiors.

Previous models have considered a normal fluid (i.e., non-superfluid)
consisting of neutrons, protons, and electrons in the core of the
neutron star. This is a valid approach if the interaction between the
different species is strong. However, theoretical calculations favor
the presence of superfluid neutrons \cite{Baym1969}. This idea is
supported by the theory of pulsar glitches \cite{Anderson1975} and by
the fact that the cooling curve of Cas\,A is consistent with a phase
transition to superfluid neutrons \cite{Shternin2011, Page2011}.
In this case the matter in the core of neutron stars cannot be
described by a single-fluid approach. The effect of superfluidity in
the oscillation spectrum of unmagnetized stars has been estimated in
\cite{Mendell1991, Mendell1998, Andersson2002, Andersson2004,
  Prix2002, Chamel2008b} and in the context of magnetars in
\cite{Glampedakis2011a, Passamonti2013, vanHoven2008, vanHoven2011,
  vanHoven2012, Andersson2009}. The main consequence of a superfluid
core is an increase in frequency of the Alfv\'en continuum bands by a
factor of several with respect to the normal fluid, for the same
magnetic field strength.  It was suggested in \cite{Passamonti2013}
that such an increase (in conjunction with stratification) could
account for the observed high frequency QPOs as fundamental, polar
($m=2$) non-axisymmetric Alfv\'en modes, although this model cannot
simultaneously accommodate the lowest observed frequency QPOs.  How
superfluid neutrons in the crust would affect the spectrum of shear
oscillations was studied both for magnetized and unmagnetized models
in~\cite{Andersson2009, Samuelsson2009, Sotani2013}.

Here, we investigate the effect of a superfluid core on the
turning-point magneto-elastic QPOs of magnetars. Superfluidity is
handled in our model by decoupling the superfluid neutrons in the core
of the neutron star completely, i.e., we assume that there is no
entrainment between neutrons and protons, and no direct interaction
between both species.  Hence, neutrons affect protons only through
their gravitational interaction. Protons are expected to be
superconducting in the core of neutron stars \cite{Baym1969}, but the
magnetic field inside a magnetar may suppress superconductivity beyond
a critical field strength that is estimated to be in the range
$10^{15}\,$G $\lesssim B_\mathrm{core}\lesssim10^{16}\,$G
\cite{Glampedakis2011a}.  Therefore, we consider normal
(non-superconducting) protons in the core. In addition, since
magnetars are slow rotators with periods of $\sim10\,$s, we neglect
effects due to rotation that could create superfluid vortices.

The results presented here are obtained with the numerical code
{\small \tt MCOCOA} that solves the general-relativistic MHD
equations~\cite{Cerda2008, Cerda2009} including a treatment of elastic
terms for the neutron star crust~\cite{Gabler2011letter, Gabler2012,
  Gabler2013}.  The influence of a superfluid phase of neutrons
coexisting with a normal fluid can be described by the entrainment, a
measure of the interaction of the different species.  In the crust the
interaction of the superfluid neutron component with the nuclei of the
lattice due to Bragg reflection is so strong \cite{Chamel2012} that
the perturbation of the lattice will carry along most of the
superfluid neutrons. Therefore, we assume complete entrainment and
treat the crust as if it was a single fluid with shear, including the
total mass of all constituents.  In the core we assume for simplicity
that the neutrons are completely decoupled, i.e., only protons are
dynamically linked to the magneto-elastic oscillations. This extreme
approximation (complete decoupling) complements the one in our
previous work, where we assumed complete coupling.
The proton fraction in the core has been estimated to be $X_p\sim0.05$
\cite{Glendenning1985, Wiringa1988, Akmal1998, Douchin2001,
  Hebeler2010}, and we have assumed this value in all our calculations
(a more detailed treatment would consider a particular
stratification).  The dynamical behavior of electrons can be neglected
because of their small mass.

For the evolution we have to solve the momentum and the induction
equation. The latter remains unchanged compared to our previous work
\cite{Cerda2009, Gabler2012}, while the former one holds now for
protons only. Effectively, we change the momentum of the fluid in the
core in the $\varphi$-direction by replacing the total rest-mass
density $\rho$ by the rest-mass density of protons $\rho_{p} = X_p
\rho$ only. The superfluid neutrons are not influenced by the
torsional magneto-elastic oscillations.

Since the system under consideration consists of crust and core that
have different properties, it is not obvious whether there exist discrete
eigenmodes. 
Hence, to differentiate between discrete and continuum oscillations
we
use the phase of the Fourier transform of the time evolution. For
discrete modes the whole star oscillates with the same phase. In
contrast, the continuum of torsional Alfv\'en oscillations gives rise
to a continuous phase change as one crosses field lines, because the
eigenfrequencies of neighboring field lines are slightly different,
i.e., the oscillation at these lines are out of phase leading to phase
mixing and damping of the oscillations.

For a magnetar model with a normal fluid core, a crust, and a poloidal
magnetic field there exist three distinct types of torsional
magneto-elastic QPOs \cite{Gabler2012}: For weak surface magnetic
fields, $B \lesssim 10^{15}\,$G, the QPOs are reflected at the
core-crust interface and the different field lines are weakly coupled
through this boundary. At strong magnetic fields, $B > 5\times
10^{15}\,$G, the field dominates over the crustal shear modulus, i.e.,
the magneto-elastic QPOs reach the surface and individual field lines
are coupled by the entire crust instead of only at the core-crust
interface.  For intermediate field strengths the magneto-elastic QPOs
change from being reflected at the core-crust interface to being
reflected at the surface of the star. In all these cases no discrete
modes exist.

\begin{figure}
\includegraphics[width=.46\textwidth]{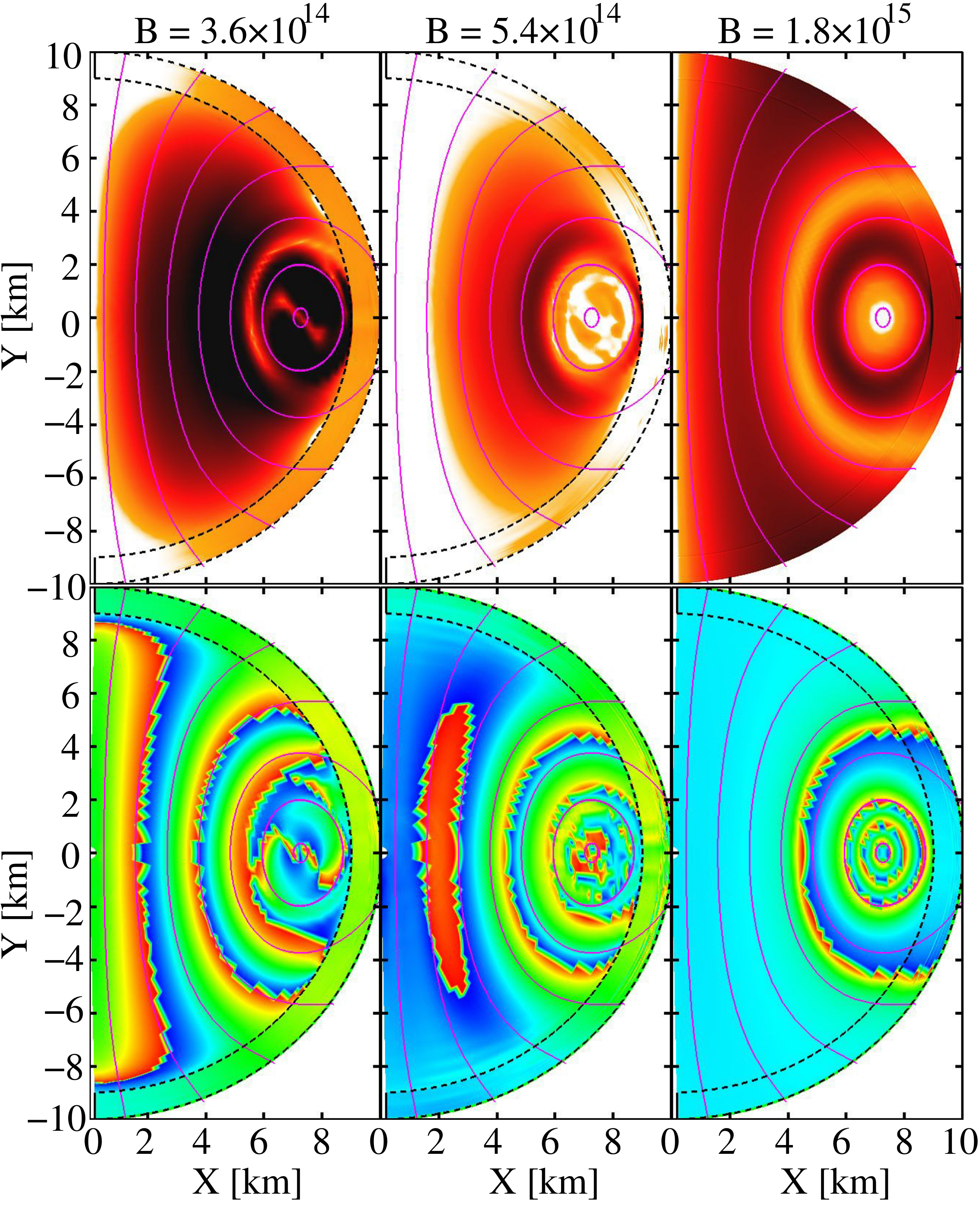}
\caption{Effective amplitude {\it top row} and phase {\it bottom row}
  of a particular QPO (see text for details) ranging from white-blue
  (minimum) to orange-red (maximum), and from $\theta=-\pi/2$ (blue)
  to $\theta=\pi/2$ (orange-red), respectively. The crust is indicated
  by the dashed black line, and magnetic field lines are given by the
  magenta lines. The bottom right panel demonstrates
  that for a typical magnetar surface field strength a discrete QPO
  with a constant phase (modulo numerical inaccuracies at a nodal
  line) exists in the whole open field line region. One can also recognize that
  the phase changes in regions with vanishing amplitudes only.}
\label{fig_modes}
\end{figure}

If all superfluid neutrons in the core are decoupled we find that
there are still no discrete modes for weak magnetic fields of $B
\lesssim \mathrm{few} \times 10^{14}\,$G. However, at typical magnetar
surface field strengths of $B \sim 10^{15}\,$G there exist QPOs with
an almost constant phase in nearly the whole open field line region.
This transition is exemplified in Fig.\,\ref{fig_modes} for the lowest
frequency QPO, but holds for other QPOs as well. The particular model
shown in Fig.\,\ref{fig_modes} has a mass of $1.4~\mathrm{M}_\odot$
and was computed with the APR EoS in the core \cite{Akmal1998} and the
DH EoS in the crust \cite{Douchin2001}.
For $B \gg 10^{15}\,$G we expect the continuum to appear again, which
is present in simulations without crust.

The different regimes can be explained with the speed of perturbations
propagating along magnetic field lines, which exhibits a discontinuity
at the core-crust interface. For weak magnetic fields $B \lesssim
10^{14}\,$G, the shear speed is much higher than
the Alfv\'en speed at the base of the crust, i.e., there is a large
jump in the propagation speed at the crust-core interface. This leads
to a significant reflection of the QPOs at the core-crust interface,
confining the QPOs mostly to the core (Fig.\ref{fig_modes}, top left
panel). In the superfluid case, the Alfv\'en speed in the core is a
factor $\sqrt{1/X_p}$ higher than for a normal fluid, because $v_A^2
\sim B^2 / \rho$ and only protons (a fraction $X_p$ of the total mass)
take part in the magneto-elastic oscillations. Hence, the jump in
propagation speed at the crust-core interface is significantly smaller
giving rise to less reflection and stronger penetration of the
magneto-elastic oscillations into the crust. At $ B \sim 10^{15}\,$G
the jump vanishes, and the Alfv\'en speed in the core approaches the
shear speed at the base of the crust. The strong coupling of
the magnetic field lines by the crust then leads to the appearance of
oscillations with constant phase in the region of open field lines.
Similar effects were observed in \cite{Levin2007} and \cite{Cerda2009}
for strong (numerical) viscosity.

The above effect is less pronounced in the normal fluid case, because
there the transition from reflection at the core-crust interface to
dominance of magnetic over shear effects in the crust occurs between
$10^{15}\,$G$\,< B < 5\times 10^{15}\,$G, while in the superfluid case
the transition already starts at a few $10^{14}\,$G. In addition, a
more massive core takes part in the magneto-elastic oscillations in
the normal fluid case, i.e., the coupling to the crust is weaker.

\begin{figure}
\includegraphics[width=.48\textwidth]{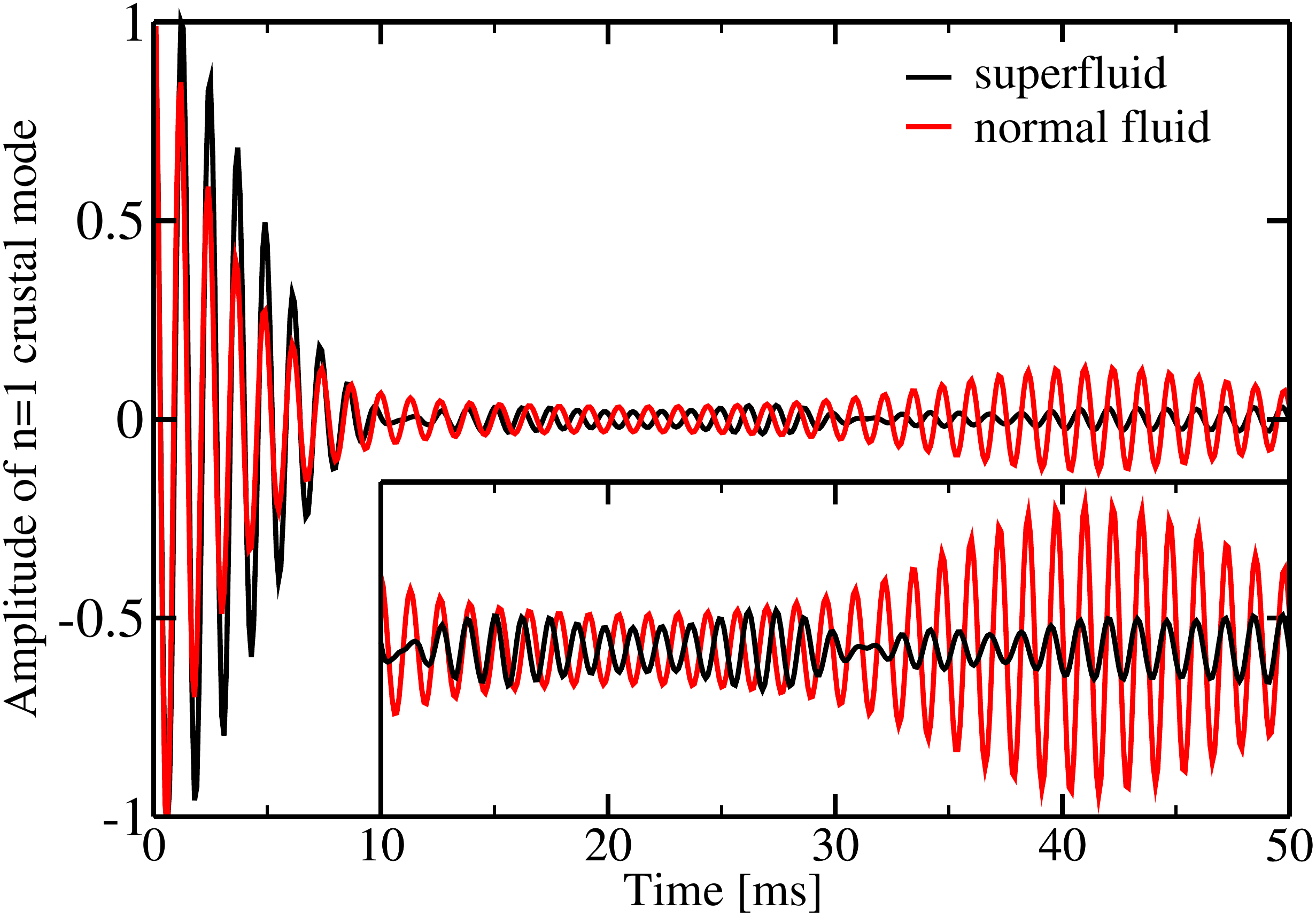}
\caption{Overlap integral with the $n=1$ crustal shear mode at
  $B=10^{15}\,$G for normal and superfluid models. The inset shows a
  magnification of the amplitude from $10$ to $50\,$ms.}
\label{fig_damp}
\end{figure}

We now turn to the high frequency QPOs with $f>500\,$Hz, whose
preferential excitation could not easily be justified in the
magneto-elastic model with a normal fluid core.  In a first attempt to
include the effects of superfluidity, VanHoven \& Levin
\cite{vanHoven2012} assumed that only 5\% of the core takes part in
the magneto-elastic oscillations. In their simulations, the $n=1$
crustal shear modes are absorbed very efficiently into the core when
initially only the crust is excited. In Fig.\,\ref{fig_damp} we show
the corresponding overlap integral (a measure for the excitation of a
given crustal mode, see \cite{Gabler2012}) for a simulation with $B =
10^{15}\,$G and with a $n=1$, $l=2$ crustal shear mode as initial
perturbation. We obtain initial damping time scales of a few
milliseconds for the superfluid and normal fluid cases, in broad
agreement with \cite{vanHoven2012}.  However, the damping does not
continue at the initial rate (see the inset in Fig.\,\ref{fig_damp}).
After about $10\,$ms almost stable oscillations with modulating
amplitudes persist at a similar frequency for both fluid models with
much lower damping rates.

\begin{figure}
\includegraphics[width=.48\textwidth]{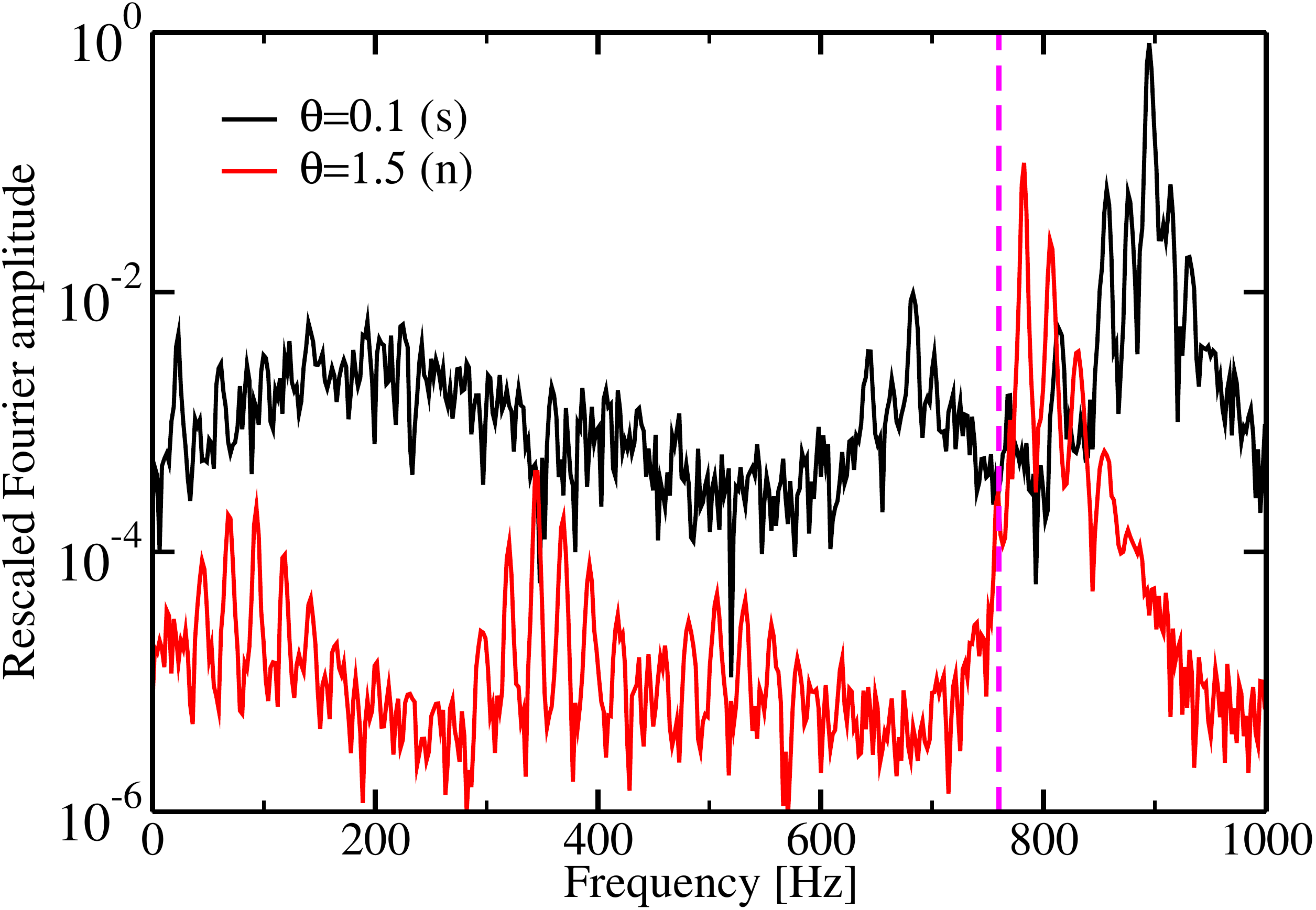}
\caption{Fourier transform of the velocity for a superfluid model near
  the polar axis ($\theta=0.1$; black line), and for a normal fluid
  model near the equator ($\theta=1.5$; red line).  The dashed magenta
  line indicates the frequency $f=760\,$Hz of the $n=1$, $l=2$ crustal
  shear mode that was used as initial perturbation.}
\label{fig_FFT}
\end{figure}

\begin{figure}
\includegraphics[width=.48\textwidth]{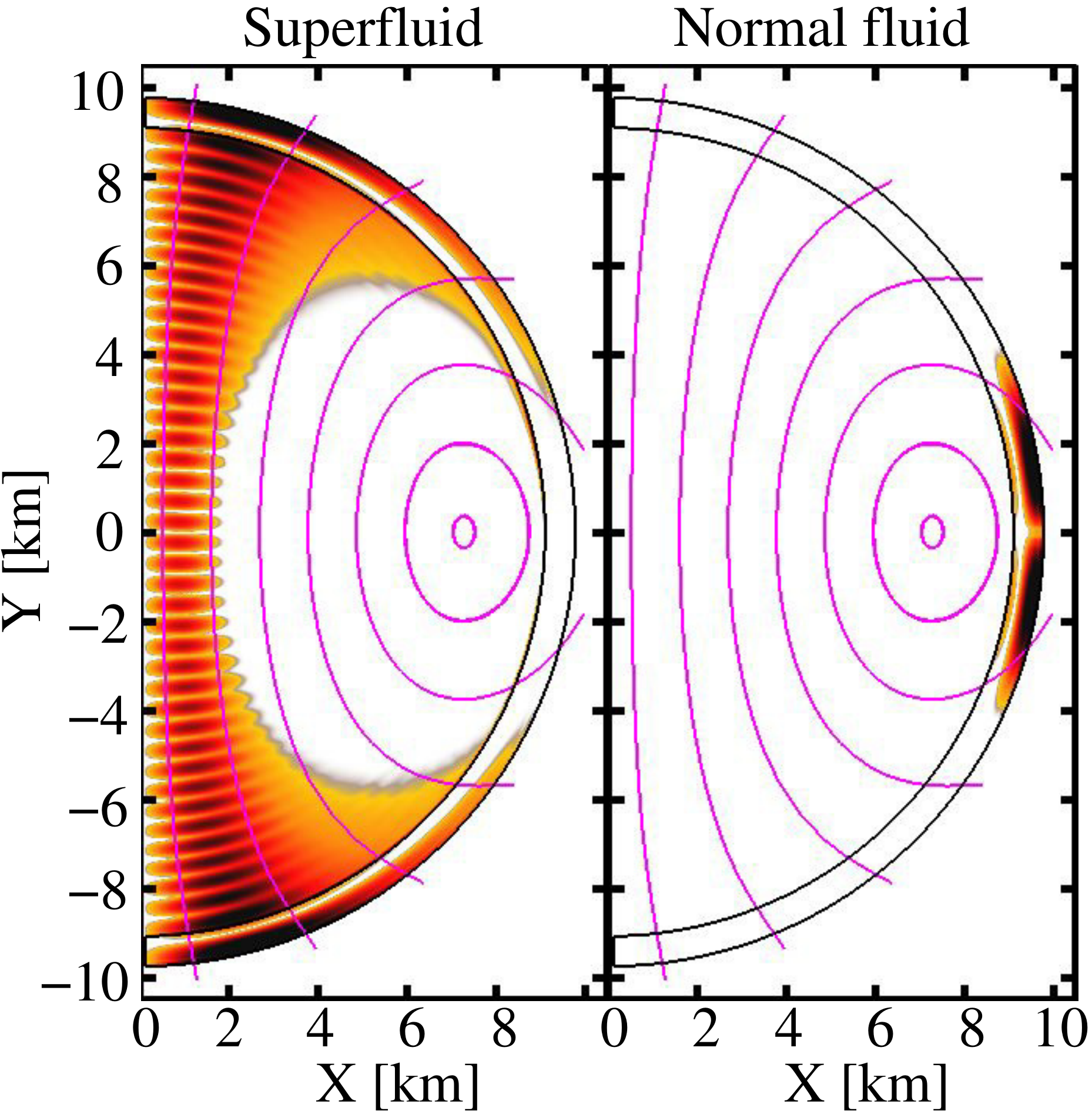}
\caption{Spatial distributions of the Fourier amplitudes of the
  velocity at the peak frequencies in Fig.\,\ref{fig_FFT}.  Left
  panel: QPO at $f \sim 893\,$Hz and $B=5.4\times 10^{14}\,$G for
  decoupled superfluid neutrons. Right panel: strongest shear
  dominated magneto-elastic ($n=1$) QPO in the crust at $f \sim
  782\,$Hz and $B=10^{15}\,$G for a normal fluid core. Magnetic field
  lines are shown by magenta lines.}
\label{fig_structure}
\end{figure}

Fourier transforming the data for evolution times of about $1\,$s we
find that the crustal $n=1$ shear mode ($f \sim 760\,$Hz) excites a
global magneto-elastic QPO with $f \sim 893\,$Hz in the superfluid
case. For the normal fluid we find three magneto-elastic QPOs in the
crust with $f \sim 782,\,806,$ and $829\,$Hz, respectively. In
Fig.\,\ref{fig_FFT} we show the Fourier amplitude of the (azimuthal)
velocity inside the crust close to the equator for the normal fluid
and close to the pole for the superfluid model. The corresponding
spatial structures of the strongest QPOs of both models are displayed
in Fig.\,\ref{fig_structure}.
In both cases the radial structure of the $n=1$ QPO remains similar to
that of the pure crustal shear mode inside the crust. However, its
angular dependence differs considerably from that of the original
spherical harmonic one due to the interaction with the core (see also
\cite{Gabler2012} for the normal fluid case).

The $n=1$ crustal shear modes propagate in radial direction. Because
the magnetic field lines are almost orthogonal to this direction close
to the equator, the coupling to the core is very weak in that
region. This explains the structure of the QPO for the normal fluid
case, together with the fact that magneto-elastic oscillations in the
core are strongly reflected at the core-crust interface, which does
not allow for a resonance between crust and core oscillations. At
stronger magnetic fields, the Alfv\'en character of the
magneto-elastic oscillations dominates, before the jump in propagation
speed at the core-crust interface disappears.
In contrast, in the superfluid case the strongest QPO at $f\sim
893\,$Hz has its maximum close to the polar axis. Here, the shear
terms dominate in the crust, and a higher magneto-elastic overtone in
the core can enter in resonance.

Overall, our results allow for a better understanding of the observed
frequencies in SGR giant flares. The inclusion of superfluidity seems
to be a key ingredient which helps in several ways: Firstly, the
observed high frequency QPOs can be explained as global
magneto-elastic QPOs resulting from a resonance between the crust and
a high ($\sim 40$) magneto-elastic overtone.  This is only possible if
there are superfluid neutrons in the core.  There still may exist
oscillations at frequencies above $500\,$Hz in models with normal
fluid cores, but since these QPOs are limited to a region close to the
equator they can only affect a small region of the magnetosphere close
to the star.  This makes it difficult to explain why QPOs are observed
at different rotational phases \cite{Strohmayer2006}.  Secondly, the
phase of the magneto-elastic QPOs becomes constant for magnetic fields
between several $10^{14}$G to several $10^{15}\,$G. Due to the absence
of phase mixing we expect that these QPOs are longer lived than
magneto-elastic QPOs of normal fluid cores. We plan to investigate
this in forthcoming work.  Thirdly, the necessary magnetic field to
match the low frequency QPOs $f \sim 30\,$Hz decreases by a factor of
$\sqrt{1/X_p}$ which reduces our previous estimates $B \sim 1 - 4
\times 10^{15}\,$G \cite{Gabler2012} to $B \sim 2\times 10^{14} -
10^{15}\,$G, in good agreement with \cite{vanHoven2008, vanHoven2012}
and current spin down estimates for magnetars showing giant flares
($6\times 10^{14} \lesssim B \lesssim 2.1\times 10^{15}\,$G).
A more realistic treatment of the entrainment is likely to further
decrease our magnetic field estimates slightly \cite{Andersson2009}.

These results do not only indicate the presence of a superfluid phase
of neutrons in the core of SGRs, but they may also constrain the EoS
of the crust significantly. The high frequency QPO and the threshold
for the outbreak of the low frequency QPOs \cite{Gabler2012} give
independent limits on the shear modulus of the crust, and hence on the
EoS. We plan to investigate this in detail in forthcoming work. For
the first time, our magnetar model that includes the effects of the
crust, the magnetic field, and superfluidity can accommodate
simultaneously all types of observed QPO frequencies, low ($f<
150\,$Hz) and high ($f> 500\,$Hz), in the giant flares of SGRs.  For a
particular model with a surface magnetic field strength of $B \approx
1.4\times 10^{15}\,$G we find low frequency oscillations at $21$,
$30$, $43$, $58$, $70$, $74$, $84$, ${89}$, $98$, $119$, $129$, $135$,
${149}$, and $162\,$Hz that are in broad agreement with the QPOs
observed in SGR 1806-20 at $18$, $26$, $30$, $92$, and $150\,$Hz.

More details of the theoretical framework and a careful analysis will
be provided in forthcoming papers.  The next major step towards a
complete model for giant flare QPOs consists in finding a modulation
mechanism of the emission in the magnetosphere.

Work supported by the Collaborative Research Center on Gravitational
Wave Astronomy of the Deutsche Forschungsgemeinschaft (DFG
SFB/Transregio 7), the Spanish {\it Ministerio de Educaci\'on y
  Ciencia} (AYA 2010-21097-C03-01) the {\it Generalitat Valenciana}
(PROMETEO-2009-103), the ERC Starting Grant CAMAP-259276, an IKY-DAAD
exchange grant (IKYDA 2012) and by CompStar, a Research Networking
Programme of the European Science Foundation. N.S. also acknowledges
support by an Excellence Grant for Basic Research (Research Committee
of the Aristotle University of Thessaloniki, 2012).  Computations were
performed at the {\it Servei d'Inform\`atica de la Universitat de
  Val\`encia}.

\bibliography{magnetar}

\end{document}